\documentclass{eptcs}

\usepackage[latin3]{inputenc}
\usepackage{amsmath}
\usepackage{amsfonts}
\usepackage{amssymb}
\usepackage{wasysym}

\usepackage{amsthm}
\newtheorem{definition}{Definition}

\newtheorem{theorem}{Theorem}

\newcommand{\bind}{>\!\!>\!\!=}
\DeclareMathOperator*{\argmax}{arg\,max}

\usepackage{haskell}
\usepackage{mathtools}
\usepackage{framed}
\usepackage{algorithm}
\usepackage{algpseudocode}

\newcommand{\K}{\mathcal K}
\newcommand{\J}{\mathcal J}
\newcommand{\concat}{+\!\!\!+}
\newcommand{\ignore}[1]{}

\title{Monad Transformers for Backtracking Search}
\author{Jules Hedges\\
Queen Mary University of London\\
\texttt{j.hedges@qmul.ac.uk}}
\def\titlerunning{Monad Transformers for Backtracking Search}
\def\authorrunning{Jules Hedges}

\begin{document}

\maketitle
\begin{abstract}
This paper extends Escard\'o and Oliva's selection monad to the selection monad transformer, a general monadic framework for expressing backtracking search algorithms in Haskell. The use of the closely related continuation monad transformer for similar purposes is also discussed, including an implementation of a DPLL-like SAT solver with no explicit recursion. Continuing a line of work exploring connections between selection functions and game theory, we use the selection monad transformer with the nondeterminism monad to obtain an intuitive notion of backward induction for a certain class of nondeterministic games.
\end{abstract}

\section{Introduction}

Selection functions are higher-order functions related to continuations, introduced by Martin Escard\'o and Paulo Oliva in \cite{escardo10a}, with many remarkable properties. There are many intuitions that can be used to understand selection functions, including viewing them as
\begin{itemize}
\item a generalised form of search algorithm
\item a generalised notion of rationality
\end{itemize}
Selection functions form a monad, called the \emph{selection monad}, which has an important intuition as a refinement of the continuation monad that carries additional information. All of these intuitions will be used in this paper. The monoidal product of the selection monad, called the \emph{product of selection functions}, can be used for several seemingly unrelated purposes:
\begin{enumerate}
\item In proof theory, it gives a computational meaning to the negative translation of the axiom of countable choice \cite{escardo10a}
\item In synthetic topology, it is a computational form of Tychonoff's theorem \cite{escardo08}
\item In functional programming, it provides a monadic framework for backtracking search problems \cite{escardo10d}
\item In game theory, it is a generalisation of the backward induction algorithm \cite{escardo12}
\end{enumerate}
In this paper we are concerned with points (3) and (4).

The intuition behind (3) is that a selection function is a generalised search algorithm, inputting a `generalised predicate' characterising objects to be found, and outputting an object that satisfies the predicate if one exists. The purpose of the product of selection functions is to combine search algorithms for simple search spaces into search algorithms for more complex search spaces. This has been used to derive so-called \emph{seemingly impossible functional programs} which search certain infinite (but topologically compact) data types in finite time \cite{escardo07}.

For (4) we have a powerful intuition that a selection function is a generalised form of preference for rational agents in game theory \cite{hedges13}. Indeed the $\argmax$ operator, which characterises the behaviour of classical economic agents, is one of the canonical examples of a selection function. The product of selection functions applied to copies of $\argmax$ is precisely the backward induction algorithm, used to compute subgame-perfect Nash equilibria of sequential games \cite{escardo12}, and we can transfer this game-theoretic intuition to other instances of the product of selection functions. An introduction to selection functions written from this point of view is \cite{escardo11}.

Part \ref{s2} of this paper is an introduction to selection functions and the Haskell implementation of the continuation and selection monads. In part \ref{s3} we show the use of the continuation monad, rather than the selection monad, for writing search algorithms in Haskell, and the use of the continuation monad transformer to express more advanced search algorithms such as DPLL. In part \ref{s4} we define the selection monad transformer and show its relationship to the ordinary selection monad and the continuation monad transformer. Part \ref{s5} defines nondeterministic sequential games, and in part \ref{s6} we use the monoidal product of the selection monad transformer, applied to the nondeterminism monad, to give an intuitive notion of backward induction for these games.

This paper uses two different notations, namely Haskell and `naive type theory', which can be read either as ordinary set theory or as functional pseudocode. To make it easy to distinguish the notations, Haskell code is set in a \fbox{box}. There are three appendices listing Haskell code verbatim: appendix \ref{dpllfull} contains the DPLL implementation (part \ref{s3}), \ref{library} contains the selection monad transformer library (part \ref{s4}) and \ref{exgame} contains an example nondeterministic game (parts \ref{s5} and \ref{s6}). All of this code can be downloaded from the author's homepage\footnote{\url{http://www.eecs.qmul.ac.uk/~julesh/}}.

\section{Quantifiers and selection functions}\label{s2}

A selection function is defined to be any function with a type of the form
\[ \varepsilon : \J_R X \]
where the selection monad $\J_R$ is defined as
\[ \J_R X = (X \to R) \to X \]
or, in Haskell notation,
\begin{framed} \begin{haskell*}
\hskwd{data} Sel r x = Sel \{runSel :: (x \to r) \to x\}
\end{haskell*} \end{framed} \noindent
There is a close relationship between the selection monad and the well-known continuation monad
\[ \K_R X = (X \to R) \to R \]
which in Haskell is
\begin{framed} \begin{haskell*}
\hskwd{data} Cont r x = Cont \{runCont :: (x \to r) \to r\}
\end{haskell*} \end{framed} \noindent
(Note that in Haskell the constructor of \<Cont\> is written \<cont\> as a consequence of the continuation monad being, in reality, defined in terms of the continuation monad transformer.) A function $\varphi : \K_R X$ is called a quantifier. Every selection function $\varepsilon : \J_R X$ induces a quantifier $\overline{\varepsilon} : \K_R X$ by
\[ \overline{\varepsilon} p = p (\varepsilon p) \]
This operation is a monad morphism from $\J_R$ to $\K_R$. When $\varphi$ is a quantifier satisfying $\varphi = \overline{\varepsilon}$ we say that $\varepsilon$ \emph{attains} $\varphi$. Not every quantifier is attainable, for example a constant quantifier $\varphi p = r_0$ is not attainable because we can find some $p : X \to R$ such that $r_0$ is not in the image of $p$. However many important quantifiers are attainable, and in those cases selection functions offer several advantages.

The canonical examples of quantifiers are the maximum operator $\max : \K_\mathbb R X$ for a finite set $X$ defined by
\[ \max p = \max_{x \in X} px \]
and the existential quantifier $\exists : \K_\mathbb B X$, where $\mathbb B = \{ \bot, \top \}$, defined by
\[ \exists p = \begin{dcases*}
\top & if $px = \top$ for some $x \in X$ \\
\bot & otherwise
\end{dcases*} \]
Both of these quantifiers are attained: $\max$ is attained by the operator $\argmax$ which produces a point at which $p$ attains its maximum, and $\exists$ is attained by Hilbert's $\varepsilon$ operator. Another interesting example is the integral operator
\[ \int : \K_\mathbb R [0,1] \]
defined by
\[ \int p = \int_0^1 px\, \mathrm d x \]
A combination of the mean value theorem and the axiom of choice proves that this quantifier is attained. This example is interesting because a computable quantifier is attained by a noncomputable selection function (the selection function $\argmax$ also has this property if we replace the finite set $X$ with a compact topological space with infinitely many points, such as the unit interval).

All monads $M$ (note the term \emph{monad} will always mean \emph{strong monad} in this paper) have a monoidal product
\[ \otimes : MX \times MY \to M(X \times Y) \]
which can be expressed in terms of unit and bind. (In fact there are always two monoidal products, but we are interested only in one of them.) A monomorphic iterated form of this product is present in the Haskell prelude as
\begin{framed}
\begin{haskell*}
sequence :: (Monad m) \implies [m a] \to m [a]
\end{haskell*}
\end{framed} \noindent
It is important to note that due to the restrictions of Haskell's type system we can only take the iterated product of selection functions and continuations which have the same type. In a dependently typed language we could express the general product, which for selection functions has type
\[ \bigotimes : \prod_i \mathcal J_R X_i \to \mathcal J_R \prod_i X_i \]

For the continuation monad the binary product is given by
\[ (\varphi \otimes \psi)q = \varphi (\lambda x^X . \psi (\lambda y^Y . q(x, y))) \]
As special cases this includes min-maxes in game theory, composition of logical quantifiers such as $\exists x^X \forall y^Y . q(x, y)$, and multiple integration over products of $[0,1]$, as well as combinations of these such as
\[ \max_{x \in X} \int_0^1 q(x, y)\, \mathrm d y \]
The product of selection functions is a more complex operation, given in \cite{escardo11} by
\[ (\varepsilon \otimes \delta)q = (a, b_a) \]
where
\begin{align*}
a &= \varepsilon (\lambda x^X . q(x, b_x)) \\
b_x &= \delta (\lambda y^Y . q(x, y))
\end{align*}
Because the overline operation is a monad morphism it commutes with the monoidal products:
\[ \overline{\varepsilon \otimes \delta} = \overline{\varepsilon} \otimes \overline{\delta} \]
In other words, if $\varepsilon$ attains $\varphi$ and $\delta$ attains $\psi$ then $\varepsilon \otimes \delta$ attains $\varphi \otimes \psi$.

One of the most important and remarkable properties of selection functions is that the product of selection functions is well-defined when iterated infinitely, so long as $R$ is compact and $q$ is continuous. Moreover the Haskell function \<sequence\> specialised to the selection monad will terminate on an infinite list, provided the Haskell datatype \<r\> is compact in the sense of \cite{escardo08}, since computable functions are continuous. Other well-known monads, including the continuation monad, do not have this property. Finite discrete types such as \<Bool\> are compact, and an example of an infinite compact type is \<Int \to Bool\>, which corresponds to the Cantor space $2^\omega$. An example of a type which is not compact is \<Int\>.

\section{SAT solving with the continuation monad transformer}\label{s3}

In this section we investigate the use of the continuation monad transformer for structuring backtracking search algorithms. (For the ordinary continuation monad this is implicit in the work of Escard\'o and Oliva.) In general, for the decision version of a search problem the continuation monad can be used. One advantage of this approach is that these monads are better-known and are part of a standard Haskell installation. A more tangible advantage is that using the continuation monad is likely to be more efficient (although the runtime of these algorithms is generally unknown). However, the main reason for this section is simply to investigate putting the continuation monad transformer to a use for which it was not originally intended.

The selection function
\begin{framed} \begin{haskell*}
\varepsilon &::& Sel Bool Bool \\
\varepsilon &=& Sel \$ {\lambda}p \to p True
\end{haskell*} \end{framed} \noindent
will solve a simple optimisation problem: given a function \fbox{\<p :: Bool \to Bool\>} the selection function $\varepsilon$ will find \<x\> making \<p x\> true, if one exists. Then
\begin{framed} \begin{haskell*}
sequence \$ repeat \varepsilon
\end{haskell*} \end{framed} \noindent
will, given a function \fbox{\<q :: [Bool] \to Bool\>}, find \<xs\> making \<q xs\> true, if one exists (recall that the Haskell function \<repeat\> builds infinite lists). In other words, this function will find satisfying assignments of propositional formulas. The correctness and totality of this function is far from obvious! The most direct proof is by bar induction (specifically, induction on the modulus of continuity of \<q\>) but a far more intuitive proof method is to view the search problem as an unbounded sequential game and apply theorem 6.2 of \cite{escardo11}.

However, the classical SAT problem is only to decide whether a satisfying assignment exists, rather than to actually compute one. By construction, if a formula $q$ has a satisfying assignment then $(\otimes_i\varepsilon) q$ is a satisfying assignment. Therefore $q$ is satisfiable iff $q ((\otimes_i \varepsilon) q)$ is true, that is, if $(\overline{\otimes_i \varepsilon}) q$ is true. If the product is finite this is equal to $(\otimes_i \exists) q$, and $\exists$ can be written directly in Haskell as
\begin{framed} \begin{haskell*}
\exists &::& Cont Bool Bool \\
\exists &=& cont \$ {\lambda}p \to p \$ p True
\end{haskell*} \end{framed} \noindent
Using \<sequence\> for the monoidal product yields an extremely small, self-contained Haskell SAT solver:
\begin{framed} \begin{haskell*}
\hskwd{import} Control.Monad.Cont \\
sat n = runCont \$ sequence \$ replicate n \$ cont \$ {\lambda}p \to p \$ p True
\end{haskell*} \end{framed}

The type is \fbox{\<sat :: Int \to ([Bool] \to Bool) \to Bool\>}, so it is not a true SAT solver in the sense that it takes its input as a function \fbox{\<[Bool] \to Bool\>} rather than in a discrete form such as a clause-set. The other input is the number of variables to search, which is necessary because \<sequence\> specialised to continuations will diverge on infinite lists. It is also important to stress that this algorithm is not a SAT solver written in continuation-passing style, rather it uses the continuation monad to directly represent the recursion.

Using the continuation monad transformer we can begin to refine this algorithm, for example we can write a DPLL-like algorithm using a state monad to store clause-sets. The DPLL algorithm, introduced in \cite{davis60}, decides the satisfiability of CNF-formulas by successively extending the formula with either a literal or its negation, and at each stage applying two simplifying transformations, namely unit clause propagation and pure literal elimination. Most modern SAT solvers are based on DPLL combined with various heuristics to improve average-case complexity, see for example \cite{silva08}.

For simplicity we implement only unit clause propagation. The algorithm we will implement is represented in imperative pseudocode in algorithm \ref{dpll}.
\begin{algorithm}
\caption{Imperative DPLL algorithm}
\label{dpll}
\begin{algorithmic}
\Function{DPLL}{$\varphi$}
	
\If{$\varphi$ is an empty clause-set} 

\Return True
\EndIf
\If{$\varphi$ contains the empty clause}

\Return False \EndIf
\While{$\varphi$ contains a unit clause $l$}

\For{each clause $c$ in $\varphi$}

\If{$c$ contains $l$}

\State $\varphi \gets$ remove $c$ from $\varphi$

\EndIf
\If{$c$ contains $\overline{l}$} 

\State $c \gets$ remove $\overline{l}$ from $c$
\EndIf
\EndFor
\EndWhile
\State $l \gets$ next literal

\Return \Call{DPLL}{$\varphi \wedge l$} $\vee$ \Call{DPLL}{$\varphi \wedge \overline{l}$}
\EndFunction
\end{algorithmic}
\end{algorithm}

We begin with a datatype representing literals:
\begin{framed} \begin{haskell*}
\hskwd{data} Literal = Positive Int | Negative Int
\end{haskell*} \end{framed} \noindent
so the type of a clause-set is \<[[Literal]]\>. The top-level function will be
\begin{framed} \begin{haskell*}
dpll &::& Int \to [[Literal]] \to Bool \\
dpll n &=& evalState s . initialState \hswhere{
s &:: State DPLL Bool \\ 
s &= runContT (sequence \$ replicate n \varphi) q \\
{} \\
\varphi &:: ContT Bool (State DPLL) Bool \\
\varphi &= ContT \$ {\lambda}p \to p True \bind p 
}
\end{haskell*} \end{framed} \noindent
Notice that in $\varphi$ the expression \fbox{\<p\ \$\ p\ True\>} is replaced by its monad transformer equivalent \fbox{\<p\ True \bind p\>}.

The type \<DPLL\> must represent the state used by the DPLL algorithm, and we need a function \fbox{\<initialState :: [[Literal]] \to DPLL\>}. Most of the actual algorithm is contained in the query function
\begin{framed} \begin{haskell*}
q :: [Bool] \to State DPLL Bool
\end{haskell*} \end{framed} \noindent
The implementation of this function is given in appendix \ref{dpllfull}. It is important to note that the monad transformer stack \fbox{\<ContT\ Bool\ (State\ DPLL)\>} will thread a single state through an entire search, however for the DPLL algorithm we want to create a new copy of the state for every recursive call. We achieve this by representing the recursion tree explicitly, making the state type \<DPLL\> a type of binary trees with leaves labelled by the necessary data. The idea is that the function \<q\> will move up the tree according to its input, interpreting \<True\> as `go left' and \<False\> as `go right'. If it finds a leaf labelled by a clause set for which satisfiability is trivial, the function returns. If not, it extends the tree according to the remaining input. A full implementation is presented in appendix \ref{dpllfull}.

As written, this program is not particularly efficient. Potentially we could use the \<IO\> monad rather than \<State\>, and produce an optimised SAT solving algorithm for example by storing the clause sets as arrays rather than lists. However optimised SAT solvers and other search algorithms require explicit control over the backtracking, which is precisely what the continuation and selection monads do not allow. In \cite{bauer12} is presented an alternative implementation of the product of selection functions that allows explicit control over backtracking. See the next section for a discussion of how \<sequence\> explores a search space.

\section{The selection monad transformer}\label{s4}

In creating a selection monad transformer, our guiding example is the generalisation in the Haskell monad transformer library from the continuation monad
\[ \K_R X = (X \to R) \to R \]
to the continuation monad transformer
\[ \K^M_R X = \K_{MR} X = (X \to MR) \to MR \]
Paralleling this, the selection monad
\[ \J_R X = (X \to R) \to X \]
is generalised to the selection monad transformer
\[ \J^M_R X = (X \to MR) \to MX \]

The Haskell code for the monad instance is \pagebreak
\begin{framed} \begin{haskell*}
\hskwd{data} SelT r m x = SelT \{runSelT :: (x \to m r) \to m x\} \\
{} \\
\hskwd{instance} (Monad m) \implies Monad (SelT r m) \hswhere{
return = SelT . const . return \\
\varepsilon \bind f = SelT \$ {\lambda}p \to \hslet{
g x = runSelT (f x) p \\
h x = g x \bind p}{
runSelT \varepsilon h \bind g}}
\end{haskell*} \end{framed} \noindent
This comes from taking the instance declaration for the ordinary selection monad and replacing certain function applications with the monadic bind of \<m\>. Similarly we obtain a monad morphism from selections to continuations by replacing a function application with monadic bind:
\begin{framed} \begin{haskell*}
toCont &::& (Monad m) \implies SelT r m x \to ContT r m x \\
toCont \varepsilon &=& ContT \$ {\lambda}p \to runSelT \varepsilon p \bind p
\end{haskell*} \end{framed} \noindent
In type-theoretic notation we continue to write this operation as an overline $\overline{\varepsilon}$. A full code listing is given in appendix \ref{library}.

The proof that the selection monad satisfies the monad laws was found using an ad-hoc computer program written by Martin Escard\'o based on an algorithm for deciding $\beta\eta$-equivalence of $\lambda$-terms, and is linked from \cite{escardo08b}. The author has verified by hand that the selection monad transformer preserves the unit laws, however the proof for the associativity law appears to be unmanageable. The author is currently working on a formally verified proof that the selection monad transformer preserves the monad laws.

A simple but extremely useful example is to use $\J^{\text{IO}}_R$ to perform a verbose backtracking search, laying bare the subtle behaviour of the product of selection functions which could previously only be investigated using \<unsafePerformIO\>. For example we can write a function 
\begin{framed} \begin{haskell*}
verboseQuery :: ([Bool] \to Bool) \to [Bool] \to IO Bool \\
\end{haskell*} \end{framed} \noindent
which will print information about the query and its result before returning. Then a verbose SAT solver is given by \pagebreak
\begin{framed} \begin{haskell*}
verboseSat &::& Int \to ([Bool] \to Bool) \to IO [Bool] \\
verboseSat n &=& f . verboseQuery \hswhere{
f :: ([Bool] \to IO Bool) \to IO [Bool] \\
f = runSel \$ sequence \$ replicate n \varepsilon \\
\varepsilon :: SelT Bool IO Bool \\
\varepsilon = SelT (\$ True)}
\end{haskell*} \end{framed} \noindent
Experiments with this function confirm that \<sequence\> will prune large parts of a search space when it is able to, but it also duplicates a certain amount of work, potentially calling \<q\> on the same input several times. It also sometimes continues to call \<q\> even after it has found a satisfying assignment (that is, after \<verboseQuery\> has printed the result \<True\>). The problem of explaining exactly how \<sequence\> explores a search space is still open, although the experimental data provided by \<verboseSAT\> and related functions may make progress possible. The interested reader is encouraged to download the code samples and experiment with this function interactively.

\section{Nondeterministic sequential games}\label{s5}

In the remaining two sections we write $X \rightrightarrows Y$ for the type of nondeterministic total functions from $X$ to $Y$. This is equal to the Kleisli arrow $X \to \mathcal P_{>0} Y$ where $\mathcal P_{>0}$ is the nondeterminism monad, whose underlying functor is the covariant nonempty powerset functor. In Haskell we replace $\mathcal P_{>0}$ with the list monad for simplicity. In order to distinguish between the use of the list monad as `poor man's nondeterminism' and its ordinary use as an ordered data structure (such as for an ordered list of moves in a game) we use a type synonym
\begin{framed} \begin{haskell*}
\hskwd{type} \mathcal{P}_{>0} a = [a]
\end{haskell*} \end{framed} \noindent
(From a software engineering point of view it would be far better to introduce $\mathcal P_{>0}$ as a \<\hskwd{newtype}\>, but using a type synonym allows us to use the usual list functions to manipulate sets without boilerplate code.)

\begin{definition}[Finite nondeterministic sequential game]
An $n$-player sequential game consists of types $X_1, \ldots, X_n$ of moves, a type $R$ of outcomes and an outcome function
\[ q : \prod_{i = 1}^n X_i \to R \]
A \emph{play} of the game is a tuple $(x_1, \ldots, x_n) \in \prod_{i = 1}^n X_i$, and $q (x_1, \ldots, x_n)$ is called the \emph{outcome} of the play.

In a nondeterministic sequential game, we instead take the outcome function to be a nondeterministic function
\[ q : \prod_{i = 1}^n X_i \rightrightarrows R \]
and we consider $q (x_1, \ldots, x_n)$ to be the set of all possible outcomes of the play $(x_1, \ldots, x_n)$.
\end{definition}

In classical game theory the outcome type will be $\mathbb R^n$, and then we interpret the $i$th element of the tuple $q(x_1, \ldots, x_n)$ to be the profit of the $i$th player.

As a running example, we will consider a simple 2-player nondeterministic game implemented in Haskell, the full code of which is presented in appendix \ref{exgame}. Each player has a choice of moves given by
\begin{framed} \begin{haskell*}
\hskwd{data} Move = Cautious | Risky
\end{haskell*} \end{framed}
The game will be zero-sum, so its outcome will be a single integer with the first player maximising and the second minimising. The nondeterministic outcome function will be given by
\begin{framed} \begin{haskell*}
\hskwd{type} Outcome = Int \\
{} \\
q &::& [Move] \to \mathcal{P}_{>0} Outcome \\
q [Cautious, Cautious] &=& [0] \\
q [Cautious, Risky] &=& [-1, 0, 1] \\
q [Risky, Cautious] &=& [-1, 0, 1] \\
q [Risky, Risky] &=& [-2, -1, 0, 1, 2] \\
\end{haskell*} \end{framed}

Consider the $i$th player of a game deciding which moves to play, having observed the previous moves $x_1, \ldots, x_{i-1}$. Assuming that the players to follow are playing according to certain constraints (such as rationality) which are common knowledge, the player can associate to each possible move $x_i$ a set of outcomes which may result after the other players' moves. Thus the player has a nondeterministic function $X_i \rightrightarrows R$, which we call the \emph{context} of player $i$'s move.

\begin{definition}[Policies of players]
An \emph{outcome policy} for player $i$ is a rule that associates each context $p : X_i \rightrightarrows R$ with a set of outcomes which the player considers to be \emph{good} following that context. Thus an outcome policy for player $i$ is a quantifier with type
\[ \varphi_i : \K^{\mathcal P_{>0}}_R X_i \]

A \emph{move policy} for player $i$ is a rule which, given a context, nondeterministically chooses a move. Thus a move policy for player $i$ is a selection function with type
\[ \varepsilon_i : \J^{\mathcal P_{>0}}_R X_i \]

A move policy $\varepsilon_i$ is called \emph{rational} for an outcome policy $\varphi_i$ if for every context $p$ and every move $x$ that may be selected by the move policy, every outcome that could result from the context given that move is a good outcome. Symbolically,
\[ \{ r \in px \mid x \in \varepsilon_i p \} \subseteq \varphi_i p \]
The left hand side of this is precisely $\overline{\varepsilon_i} p$, where $\overline{\ \cdot\ }$ is the monad morphism $\mathcal J^{\mathcal P_{>0}}_R \to \mathcal K^{\mathcal P_{>0}}_R$. (However the subset relation prevents us from obtaining a game-theoretic interpretation of more general instances of the selection monad transformer.)

An outcome policy is called \emph{realistic} iff, for every context, there is a move such that every possible outcome of that move in that context is a good outcome. In other words, an outcome policy is realistic iff it has a rational move policy.
\end{definition}

It was pointed out by an anonymous referee that this definition of realistic is equivalent, using the axiom of choice, to the existence of an ordinary (deterministic) selection function $\varepsilon_i : (X \to \mathcal P_{>0} R) \to X$ with the property that $p (\varepsilon_i p) \subseteq \varphi_i p$. It is then possible to replace $\mathcal P_{>0} R$ with an arbitrary poset and continue using only ordinary quantifiers and selection functions. However, the point intuitively is that since we are working with nondeterministic games it makes more sense to allow the players to choose moves nondeterministically, rather than using the axiom of choice to `determinise' the move choices. More concretely, it should be possible to construct outcome policies with a computable rational move policy but no computable selection function with this property, at least if we use true nondeterminism rather than approximating it with lists.

For deterministic games there are canonical realistic outcome policies in the case when the outcome type is $\mathbb R^n$, namely maximisation with respect to the ordering of $\mathbb R$:
\[ \varphi_i p = \max_{x \in X_i} (px)_i \]
where $(px)_i$ is the $i$th coordinate projection of the vector $px : \mathbb R^n$. However for nondeterministic games there is no canonical policy, even when the outcome type is $\mathbb R$: a player might always choose $x$ such that $px$ contains the largest possible element, or they might have a more complicated policy to mitigate risk, for example preferring possible outcomes $\{ 5 \}$ to $\{ 10, -100 \}$. We leave this undetermined and allow the policies to be arbitrary.

We will consider two pairs of move policies for the example game, also called \<cautious\> and \<risky\>. For the implementation of these we begin with the type
\begin{framed}\begin{haskell*}
\hskwd{type} Choice &=& \mathcal{P}_{>0} (Move, \mathcal{P}_{>0} Outcome) \to (Move, \mathcal{P}_{>0} Outcome) \\
\end{haskell*} \end{framed} \noindent
representing `choice functions' which, given a set of move-outcome pairs, will choose one. Next we write a function called \<argopt\>, which is a generic way to convert a choice function into a selection function.
\begin{framed} \begin{haskell*}
argopt &::& Choice \to SelT Outcome \mathcal{P}_{>0} Move \\
argopt f &=& SelT \$ {\lambda}p \to [fst \$ f [(Cautious, p Cautious), (Risky, p Risky)]]
\end{haskell*} \end{framed} \noindent
It should be noted that Haskell does not allow a type synonym with a free type variable to be used as a parameter, so \fbox{\<SelT\ Outcome\ \mathcal{P}_{>0}\ Move\>} must be written \fbox{\<SelT\ Outcome\ []\ Move\>} in real code.

Now we implement our two pairs of choice functions:
\begin{framed} \begin{haskell*}
riskymax, riskymin, cautiousmax, cautiousmin &::& Choice \\
riskymax &=& maximumBy \$ comparing \$ maximum . snd \\
riskymin &=& minimumBy \$ comparing \$ minimum . snd \\
cautiousmax &=& riskymax . filter (all (\geq (-1)) . snd) \\
cautiousmin &=& riskymin . filter (all (\leq 1) . snd)
\end{haskell*} \end{framed} \noindent
In words, the risky move policies will select moves to \emph{maximise} the player's \emph{maximum} possible gain (which for the second player is \emph{minimising} the \emph{minimum} outcome), ignoring the risk of bad outcomes. The cautious move policies do likewise, but avoid moves that lead to the possibility of the worst possible outcome, which is $\pm 2$. The risky move policies are deterministic in the sense that they always return a list of length 1. The cautious move policies also have this property for the particular game we are considering, but not in general (indeed, as written the functions \<cautiousmax\> and \<cautiousmin\> are partial functions: if the \<filter\> returns an empty list then an exception will be thrown). 

\begin{definition}[Strategy]
A strategy for player $i$ in a game is a function
\[ \sigma_i : \prod_{j = 1}^{i-1} X_j \to X_i \]
which chooses a move given the observed previous moves. For a nondeterministic sequential game we take the strategies to be nondeterministic functions
\[ \sigma_i : \prod_{j = 1}^{i-1} X_j \rightrightarrows X_i \]
Thus in particular a strategy for player 1 is $\sigma_1 : \mathcal P_{>0} X_1$. A tuple $\sigma$ containing a strategy for each player is called a \emph{strategy profile}.
\end{definition}

We can understand this definition intuitively by imagining that nondeterminism is resolved at several points during the course of playing a game. First, player 1 nondeterministically chooses a move. Then the nondeterminism resolves, and some concrete move is played. Player 2 observes the actual move that was played, and nondeterministically chooses a move, and so on. Eventually, after all players have moved, we obtain a concrete play $(x_1, \ldots, x_n)$. Then the rules of the game nondeterministically determine an outcome, and finally this nondeterminism resolves to produce the actual outcome.

\section{Backward induction for nondeterministic games}\label{s6}

Now we need to define what it means for a strategy profile to be optimal. In game-theoretical language an optimal strategy profile is called a \emph{subgame-perfect Nash equilibrium}.

Intuitively, a strategy profile is optimal if for all players $i$ and all partial plays up to player $i-1$, every outcome that could result from playing the strategy profile is a good outcome in the context which, following player $i$'s move, the remaining players play according to the strategy profile.

We spell out this definition explicitly. Given a partial play $\vec x = (x_1, \ldots, x_{i-1})$, the set all moves that might be made by player $i \leq j \leq n$ by playing the strategy profile $\sigma$ is given inductively by
\[ b^{\vec x}_j = \{ x \in \sigma_j (\vec x, x_i, \ldots, x_{j-1}) \mid x_k \in b^{\vec x}_k,\ i \leq k < j \} \]
which in monad notation is
\[ b^{\vec x}_j = \left( \bigotimes_{k=i}^{j-1} b^{\vec x}_k \right) \bind \sigma_j \vec x \]
where $\otimes$ is the monoidal product of $M$, which for $M = \mathcal P_{>0}$ is the cartesian product. The set of all outcomes that might result from playing the strategies $\sigma$ on the partial play $\vec x$ is 
\[ \{ r \in q(\vec x, x_i, \ldots, x_n) \mid x_k \in b^{\vec x}_k,\ i \leq k \leq n \} \]
Now we need to define the context in which player $i$ chooses a move. Suppose after the partial play $\vec x$, player $i$ chooses the move $x \in X_i$. The set of outcomes that could result from this is simply the result of the previous calculation applied to the partial play $(\vec x, x)$, namely
\[ \{ r \in q(\vec x, x, x_{i+1}, \ldots, x_n) \mid x_k \in b^{\vec x, x}_k,\ i < k \leq n \} \]

\begin{definition}[Optimal strategy profile]
A strategy profile $\sigma$ is called optimal for given outcome policies $\varphi_i$ if for all partial plays $\vec x = (x_1, \ldots, x_{i-1})$ for $1 \leq i \leq n$ we have
\[ \{ r \in q(\vec x, x_i, \ldots, x_n) \mid x_k \in b^{\vec x}_k,\ i \leq k \leq n \} \subseteq \varphi_i p \]
where the context $p : X_i \rightrightarrows R$ is defined by
\[ px = \{ r \in q(\vec x, x, x_{i+1}, \ldots, x_n) \mid x_k \in b^{\vec x, x}_k,\ i < k \leq n \} \]
\end{definition}

Notice that an optimal strategy profile guarantees good outcomes for all players, despite the nondeterminism. In order for this to hold there must be enough good outcomes, for which it is sufficient that the outcome policies are realistic.

\begin{theorem}
A nondeterministic sequential game in which all players' outcome policies are realistic has an optimal strategy profile.
\end{theorem}

If the players' move policies are $\varepsilon_i$ and the outcome function is $q$ then an optimal strategy profile is given by
\[ \sigma_i \vec x = \pi_i \left( \left( \bigotimes_{j=i}^n \varepsilon_j \right) (q_{\vec x}) \right) \]
where $\otimes$ is the monoidal product in the monad $\J^{\mathcal P_{>0}}_R$, $q_{\vec x}$ is the partial application of $q$ to $\vec{x}$, and $\pi_i$ is the projection onto $X_i$. These can be computed in Haskell by
\begin{framed} \begin{haskell*}
strategy es q xs = head \$ runSelT \varepsilon (q . (xs \concat)) \hswhere{
\varepsilon = sequence \$ drop (length xs) es}
\end{haskell*} \end{framed} \noindent
Moreover the set of all plays which may occur from playing this strategy profile is given by
\[ \left( \bigotimes_{i=1}^n \varepsilon_i \right) q \]
which is computed in list form by the program
\begin{framed} \begin{haskell*}
plays = runSelT . sequence
\end{haskell*} \end{framed}

The corresponding expression using the monoidal product in $\J_R$ was shown in \cite{escardo12} to be a generalisation of backward induction, a standard and intuitive algorithm in classical game theory to find equilibria of sequential games. By taking the monoidal product in $\J^{\mathcal P_{>0}}_R$, we also gain a generalisation of backward induction to nondeterministic games.

It was remarked above that the appearance of the subset relation in the definition of rational move policies and optimal strategy profiles prevents us from giving these definitions for an arbitrary monad $M$. The definition given applies to submonads of the nondeterminism monad, which include finite nondeterminism (finite lists), the exception (\<Maybe\>) monad and the identity monad (in which case we get Escard\'o and Oliva's definitions). It might be possible to interpret the subset relation for some other monads in \emph{ad hoc} ways, for example as a refinement relation for the \<IO\> monad, which should have a game-theoretic meaning as games which interact with an environment outside of the game. However, the given expressions and Haskell functions to compute optimal strategy profiles works for any $M$, so we can say that this strategy profile is always optimal, despite having no general definition. It remains to be seen what game-theoretic intuition exists for the many different monads that are available.

We need to find an explicit expression for the monoidal product
\[ \otimes : \J^M_R X \times \J^M_R Y \to \J^M_R (X \times Y) \]
Starting with the usual expression for the product of selection functions
\[ (\varepsilon \otimes \delta)q = (a, b_a) \]
where
\begin{align*}
a &= \varepsilon (\lambda x^X . q(x, b_x)) \\
b_x &= \delta (\lambda y^Y . q(x, y))
\end{align*}
we replace a function application with monadic bind, and the cartesian pairing with the monoidal product $\otimes$ of $M$, yielding
\[ (\varepsilon \otimes \delta) q = a \otimes (a \bind \lambda x^X . b_x) \]
where
\begin{align*}
a &= \varepsilon (\lambda x^X . (b_x \bind \lambda y^Y . q(x, y))) \\
b_x &= \delta (\lambda y^Y . q(x, y))
\end{align*}
It will be helpful to write $q$ in a curried form, in which case we have
\begin{align*}
a &= \varepsilon (\lambda x^X . (b_x \bind qx)) \\
b_x &= \delta (qx)
\end{align*}

We prove the theorem in the simple case of a 2-move sequential game. Let the move types be $X$ and $Y$, so the outcome function has type $q : X \times Y \rightrightarrows R$. Let the outcome policies be $\varphi : \K^{\mathcal P_{>0}}_R X$ and $\psi : \K^{\mathcal P_{>0}}_R Y$, and the rational move policies be $\varepsilon : \J^{\mathcal P_{>0}}_R X$ and $\delta : \J^{\mathcal P_{>0}}_R Y$. We need to prove that the strategy profile
\begin{align*}
\sigma_X &= \{ x \in X \mid (x, y) \in (\varepsilon \otimes \delta) q \} \\
\sigma_Y x &= \delta (qx)
\end{align*}
is optimal. Using the explicit expression for the monoidal product we have
\[ (\varepsilon \otimes \delta)q = \{ (x, y) \in X \times Y \mid x \in a, y \in b_x \} \]
where
\begin{align*}
a &= \varepsilon \left( \lambda x . \{ r \in q(x, y) \mid y \in b_x \} \right) \\
b_x &= \delta (qx)
\end{align*}
therefore the first player's strategy is given directly by
\[ \sigma_X = a \]

We can directly calculate
\begin{align*}
b^{()}_X &= a \\
b^{()}_Y &= \{ y \in \delta (qx) \mid x \in a \} \\
b^{(x)}_Y &= \delta (qx)
\end{align*}
The conditions we need to verify are
\begin{enumerate}
\item $\{ r \in q(x,y) \mid x \in a, y \in \delta(qx) \} \subseteq \varphi (\lambda x . \{ r \in q(x, y) \mid y \in \delta(qx) \})$
\item For all $x : X$, $\{ r \in q(x, y) \mid y \in \delta(qx) \} \subseteq \psi(qx)$
\end{enumerate}
Since $\delta$ is a rational move policy for $\psi$ we have $\{ r \in py \mid y \in \delta p \} \subseteq \psi p$ for all contexts $p : Y \rightrightarrows R$. Using the context $py = q(x, y)$ gives the second condition. For the first condition we take the context
\[ px = \{ r \in q(x, y) \mid y \in \delta(qx) \} \]
so that $a = \varepsilon p$. Then we have
\begin{align*}
&\{ r \in q(x, y) \mid x \in a, y \in \delta (qx) \} \\
=\ &\{ r \in px \mid x \in \varepsilon p \} \\
\subseteq\ &\varphi p \\
=\ &\varphi (\lambda x . \{ r \in q(x, y) \mid y \in \delta(qx) \})
\end{align*}
This completes the proof.

For the example game the optimal plays can be computed interactively:
\begin{framed} \begin{haskell*}
&& plays [argopt cautiousmax, argopt cautiousmin] q \\
&\implies& [[Risky, Cautious]] \\
&& plays [argopt cautiousmax, argopt riskymin] q \\
&\implies& [[Cautious, Risky]] \\
&& plays [argopt riskymax, argopt cautiousmin] q \\
&\implies& [[Risky, Cautious]] \\
&& plays [argopt riskymax, argopt riskymin] q \\
&\implies& [[Risky, Risky]]
\end{haskell*} \end{framed} \noindent
Thus, each combination of cautious and risky move policies results in a different deterministic play. The intuition behind these results is that the `personality' defined by a move policy is common knowledge of the players. The only possibly unexpected result is the first, when both players are cautious. In this case the first player plays risky because she knows that the second player will avoid the maximum risk, meaning the risky move is safe.

\bigskip \textbf{Acknowledgement} The author gratefully acknowledges EPSRC grant EP/K50290X/1 which funded this work.

\appendix
\renewcommand\thesection{Appendix \Alph{section}}
\section{DPLL implementation} \label{dpllfull}

\begin{verbatim}
import Data.List
import Control.Monad.State
import Control.Monad.Trans.Cont

-- Literals

data Literal = Positive Int | Negative Int deriving (Show, Eq)

negateLiteral :: Literal -> Literal
negateLiteral (Positive n) = Negative n
negateLiteral (Negative n) = Positive n

-- Internal state of algorithm

data DPLL = Leaf {simplified :: Bool, clauses :: [[Literal]]}
          | Node {trueBranch :: DPLL, falseBranch :: DPLL}

-- Top level of algorithm

dpll :: Int -> [[Literal]] -> Bool
dpll n = evalState s . initialState where
     s = runContT (sequence $ replicate n e) q

e :: ContT Bool (State DPLL) Bool
e = ContT $ \p -> p True >>= p 

q :: [Bool] -> State DPLL Bool
q bs = do s <- get
          let (s', b) = queryState s bs 0
          put s'
          return b

-- Interactions with internal state

initialState :: [[Literal]] -> DPLL
initialState cs = Leaf {simplified = False, clauses = cs}

queryState :: DPLL -> [Bool] -> Int -> (DPLL, Bool)
queryState (Leaf False cs) bs n = queryState (Leaf True $ simplify cs) bs n
queryState s@(Leaf True []) _ _ = (s, True)
queryState s@(Leaf True [[]]) _ _ = (s, False)
queryState (Leaf True cs) bs n = queryState (Node l r) bs n where
           l = Leaf False ([Negative n] : cs)
           r = Leaf False ([Positive n] : cs)
queryState (Node l r) (True : bs) n = (Node l' r, b) where
           (l', b) = queryState l bs (n + 1)
queryState (Node l r) (False : bs) n = (Node l r', b) where
           (r', b) = queryState r bs (n + 1)

-- Operations on clause sets

simplify :: [[Literal]] -> [[Literal]]
simplify cs = if null units
              then cs
              else simplify (foldl (flip propagateUnit) cs' units) where
         (units, cs') = partition isUnitClause cs

isUnitClause :: [Literal] -> Bool
isUnitClause [_] = True
isUnitClause _ = False

propagateUnit :: [Literal] -> [[Literal]] -> [[Literal]]
propagateUnit [l] = (map $ filter (/= negateLiteral l))
                  . (filter $ notElem l)
\end{verbatim}

\section{The selection monad transformer library}\label{library}

\begin{verbatim}
-- Selection monad transformer library
-- A generic backtracking search and auto-pruning library

module SelT (SelT(..), Sel, toCont,
             boundedBinarySearch, unboundedBinarySearch) where

import Control.Monad.Cont
import Data.Functor.Identity

-- Selection monad transformer

newtype SelT r m x = SelT {runSelT :: (x -> m r) -> m x}

instance (Monad m) => Monad (SelT r m) where
         return = SelT . const . return
         e >>= f = SelT $ \p -> let g x = runSelT (f x) p
                                    h x = g x >>= p
                                 in runSelT e h >>= g

instance (MonadIO m) => MonadIO (SelT r m) where
         liftIO = lift . liftIO

instance MonadTrans (SelT r) where
         lift = SelT . const

-- Monad morphism from selections to continuations

toCont :: (Monad m) => SelT r m x -> ContT r m x
toCont e = ContT $ \p -> runSelT e p >>= p

-- Vanilla selection monad

type Sel r = SelT r Identity

-- Generic search functions

unboundedBinarySearch :: (Monad m) => SelT Bool m [Bool]
unboundedBinarySearch = sequence $ repeat $ SelT ($ True)

boundedBinarySearch :: (Monad m) => Int -> SelT Bool m [Bool]
boundedBinarySearch n = sequence $ replicate n $ SelT ($ True)
\end{verbatim}

\section{Example game}\label{exgame}

\begin{verbatim}
import SelT
import Data.List (maximumBy, minimumBy)
import Data.Ord (comparing)

data Move    = Cautious | Risky deriving (Show)
type Outcome = Int
type P a = [a]

q :: [Move] -> P Outcome
q [Cautious, Cautious] = [0]
q [Cautious, Risky]    = [-1, 0, 1]
q [Risky, Cautious]    = [-1, 0, 1]
q [Risky, Risky]       = [-2, -1, 0, 1, 2]

type Choice = P (Move, P Outcome) -> (Move, P Outcome)

argopt :: Choice -> SelT Outcome [] Move
argopt f = SelT $ \p -> [fst $ f [(Cautious, p Cautious), (Risky, p Risky)]]

riskymax, riskymin, cautiousmax, cautiousmin :: Choice
riskymax = maximumBy $ comparing $ maximum . snd
riskymin = minimumBy $ comparing $ minimum . snd
cautiousmax = riskymax . filter (all (>= (-1)) . snd)
cautiousmin = riskymin . filter (all (<= 1) . snd)

es1, es2, es3, es4 :: [SelT Outcome [] Move]
es1 = [argopt riskymax, argopt riskymin]
es2 = [argopt riskymax, argopt cautiousmin]
es3 = [argopt cautiousmax, argopt riskymin]
es4 = [argopt cautiousmax, argopt cautiousmin]

solve :: [SelT Outcome [] Move] -> P [Move]
solve es = runSelT (sequence es) q
\end{verbatim}

\bibliographystyle{eptcs}
\bibliography{/Users/jules/Dropbox/Work/refs}

\end{document}